\documentclass[conference,onecolumn]{IEEEtran}

\IEEEoverridecommandlockouts                              

\usepackage{amsmath}
\usepackage{amssymb}
\usepackage{amsthm}
\usepackage{hyperref}
\usepackage{graphicx}
\usepackage{enumerate}
\usepackage{caption}
\usepackage{subcaption}
\usepackage{color}
\usepackage{comment}
\usepackage{epstopdf}
\usepackage{float}

\def\cX{\mathcal{X}}
\def\cY{\mathcal{Y}}

\def\cP{\mathcal{P}}

\def\cS{\mathcal{S}}
\def\cF{\mathcal{F}}

\newtheorem{lemma}{Lemma}

\newtheorem{conjecture}{Conjecture}

\newcommand{\ve}[1]{\underline{#1}}
\newcommand{\eqdef}{\stackrel{\scriptscriptstyle \triangle}{=}}

\newcommand{\EE}{\mathbb{E}}

\title{Variable-length codes for channels with memory and feedback: error-exponent lower bounds}

\author{Achilleas Anastasopoulos and Jui Wu
	\thanks{The authors are with the Department
	of Electrical Engineering and Computer Science, University of Michigan, Ann
	Arbor, MI, 48105 USA e-mail: { \{anastas, juiwu\}@umich.edu}}
	}

\begin{document}
\maketitle

\begin{abstract}
The reliability function of memoryless channels with noiseless feedback and variable-length coding
has been found to be a linear function of the average rate in the classic work of Burnashev.
In this work we consider unifilar channels with noiseless feedback and study specific transmission schemes, the performance of which provides lower bounds for the channel reliability function. In unifilar channels the channel state evolves in a deterministic fashion based on the previous state, input, and output, and is known to the transmitter but is unknown to the receiver.
We consider a two-stage transmission scheme.
In the first stage, both transmitter and receiver summarize their common information in an M-dimensional vector with elements in the state space of the unifilar channel and an M-dimensional probability mass function, with M being the number of messages.
The second stage, which is entered when one of the messages is sufficiently reliable, is resolving a binary hypothesis testing problem.
The analysis assumes the presence of some common randomness shared by the transmitter and receiver, and is based on the study of the log-likelihood ratio of the transmitted message posterior belief, and in particular on the study of its multi-step drift.
Simulation results confirm that the bounds are tight compared to the upper bounds derived in a companion paper.
\end{abstract}

\section{Introduction}
\label{sec:intro}

There exists a substantial body of literature on transmission schemes for memoryless channels with noiseless feedback.
Horstein~\cite{Ho63} proposed a simple sequential transmission scheme which is capacity-achieving and provides larger error exponents than traditional fixed-length block-coding for discrete memoryless channels (DMCs).
Similarly, Schalkwijk and Kailath~\cite{ScKa66} showed that capacity and a double exponentially decreasing error probability can be achieved by a simple sequential transmission scheme for the additive white Gaussian noise (AWGN) channel  with average power constraint.
A remarkable result was derived by Burnashev in~\cite{Bu76}, where error exponent matching upper and lower bounds were derived for DMCs with feedback and  variable-length codes. The error exponent has a simple form $E(\overline{R})=C_1(1-\overline{R}/C)$, where $\overline{R}$ is the average rate, $C$ is the channel capacity and $C_1$ is the maximum divergence that can be obtained in the channel for a binary hypothesis testing problem.
Recently, Shayevitz and Feder~\cite{ShFe11} identified an underlying principle shared by the aforementioned schemes and introduced a simple encoding scheme, namely the posterior matching  (PM) scheme for general memoryless channels and showed that it achieves capacity.
The above transmission schemes can be contrasted to those inspired by the work in~\cite{YaIt79}
where a variable-length transmission scheme was proposed and its error exponent was found to achieve the Burnashev upper bound. This scheme (and others inspired by it) is not explicit in the sense that it assumes that some unspecified capacity-achieving codes are used in the ``communication'' stage of the transmission.

For channels with memory and noiseless feedback, there exists a rich literature on the capacity characterization~\cite{viswanathan1999capacity, chen2005capacity, yang2005feedback, PeCuVaWe08, TaMi09}.
Recently, the capacity of the trapdoor channel was found in closed form in~\cite{PeCuVaWe08}, and extended to a subset of chemical channels in~\cite{WuAn16b}, while the capacity of the
binary unit memory channel on the output was found in closed form in~\cite{StChKu16}.
A number of ``explicit'' transmission schemes have been recently studied in the literature~\cite{BaAn10,An12a} but no results on error exponents are reported.
In the case of channels with memory and feedback, an error exponent analysis is performed in~\cite{TaMi09} for fixed length coding.
The only work that studies error exponents for variable-length codes for channels with memory and feedback is~\cite{CoYuTa09} where the authors consider a finite state channel with channel state known causally to both the transmitter and the receiver. The transmission scheme presented therein is inspired by that of~\cite{YaIt79} and as a result it is based on an otherwise unspecified capacity-achieving code for this channel.

In this work, we consider channels with memory and feedback, and propose and analyze variable-length transmission schemes. We specifically look at unifilar channels since for this family, the capacity has been characterized in an elegant way through the use of Markov decision processes (MDPs)~\cite{PeCuVaWe08}.
We consider a two-stage sequential transmission scheme, similar to the one proposed in~\cite{Bu76}.
In the first stage, the encoding is a time-invariant function that depends on a summary of the available common information between the transmitter and the receiver in the form of two $M$-dimensional vectors: one is the vector of current states conditioned on each message and the other is the posterior probability mass function of the message given the observation (with $M$ being the number of messages).
The second stage, which is entered when one of the messages is sufficiently reliable, is resolving a binary hypothesis testing problem much like the original scheme of Burnashev. Following the hints from our error exponent upper bound analysis in a companion paper~\cite{AnWu17b}, the second stage employs a more sophisticated transmission scheme (compared to that for DMCs) in order to achieve the error-exponent upper bound.
The analysis assumes the presence of some common randomness shared by the transmitter and receiver\footnote{It is interesting to note that the analysis of Burnashev in~\cite{Bu76} also assumes that common randomness is present although this is not explicitly stated.}, and is based on the study of the log-likelihood ratio of the transmitted message posterior belief, and in particular on the study of its multi-step drift.
Our final result is in the form of a conjecture since there is a currently unresolved issue in the analysis. However, we provide strong evidence supporting this conjecture through numerical evaluations.
Simulation results for the trapdoor, chemical, and other binary input/state/output unifilar channels
confirm that the bounds are tight compared to the upper bounds derived in~\cite{AnWu17b}.

The main difference between our work and that in~\cite{CoYuTa09} is that for unfilar channels, the channel state is not observed at the receiver. In addition, in this work, an ``explicit'' transmission scheme is proposed and analyzed.

The remaining part of this paper is organized as follows.
In section~\ref{sec:model}, we describe the channel model for the unifilar channel and its capacity characterization. In section~\ref{sec:transmission}, we propose a two-stage transmission scheme with common randomness. In section~\ref{sec:analysis}, we analyze the proposed scheme.
Section~\ref{sec:example} presents numerical evidence for the performance of the proposed schemes for several unifilar channels. Final conclusions are given in section~\ref{sec:conclusions}.

\section{Channel Model and preliminaries}
\label{sec:model}

Consider a family of finite-state point-to-point channels with inputs $X_t\in\cX \eqdef\{0,1,\ldots,|\cX|-1\}$, output $Y_t\in\cY$ and state $S_t\in\cS$ at time $t$, with all alphabets being finite and the initial state $S_1$ known to both the transmitter and the receiver.
The channel conditional probability is
\begin{align}
&P(Y_t,S_{t+1}|X^t, Y^{t-1}, S^t) = Q(Y_t|X_t, S_t) \delta_{g(S_t,X_t,Y_t)}(S_{t+1}),
\end{align}
for a given stochastic kernel $Q\in \cX\times\cS\rightarrow \cP(\cY)$ and deterministic function $g\in \cS\times\cX \times\cY\rightarrow \cS$, where $\cP(\cY)$ denotes the space of all probability measure on $\cY$, and $\delta_{a}(b)$ is the indicator function of the event $a=b$.

This family of channels is referred to as unifilar channels~\cite{PeCuVaWe08}. The authors in~\cite{PeCuVaWe08} have derived the capacity $C$ in the form of
\begin{equation}
C =  \lim_{N\rightarrow \infty}\sup_{\{p(x_t|s_t,y^{t-1},s_1)\}_{t\geq 1}}\frac{1}{N}\sum_{i=1}^{N} I(X_t,S_t;Y_t|Y^{t-1},S_1). \label{eq:capacity}
\end{equation}
The capacity can be written as an optimal reward per unit time of an appropriately defined MDP~\cite{PeCuVaWe08, WuAn16b}. For channels with ergodic behavior, the capacity has a single-letter expression
\begin{equation}\label{eq:cap_singleletter}
C = \sup_{P_{X|SB}} I(X_t,S_t; Y_t |B_{t-1}),
\end{equation}
where $B_{t-1}\in \cP(\cS)$ is the posterior belief on the current state $S_t$ given $(Y^{t-1},S_1)$ at time $t$  and the mutual information is evaluated using the distribution
\begin{align}
P&(X_t,S_t,Y_t,B_{t-1}) \nonumber \\
 &=  Q(Y_t|X_t,S_t)P_{X|SB}(X_t|S_t,B_{t-1})B_{t-1}(S_t)\pi_B (B_{t-1}),
\end{align}
where $\pi_B$ is the stationary distribution of the Markov chain $\{B_{t-1}\}$ with transition kernel
\begin{align} \label{eq:B_MC}
P_{B|B}&(B_t|B_{t-1}) = \sum_y \delta_{\phi(B_{t-1},y)}(B_t) P_{Y|B}(y|B_{t-1}),
\end{align}
where $P_{Y|B}(y|B_{t-1})=\sum_{x,s} Q(y|x,s)P_{X|SB}(x|s,B_{t-1})B_{t-1}(s)$ and the update function $B_t=\phi(B_{t-1},Y_t)$ is defined through the recursion
\begin{subequations}
\begin{align}\label{eq:B_rec}
&\quad B_t(s) =\nonumber \\
&\frac{\sum_{x,\tilde{s}}\delta_{g(\tilde{s},x,Y_t)}(s)Q(Y_t|x,\tilde{s})P_{X|SB}(x|\tilde{s},B_{t-1})B_{t-1}(\tilde{s})}
       {P_{Y|B}(Y_t|B_{t-1})}.
\end{align}
\end{subequations}
In this paper, we restrict our attention to such channels with strictly positive $Q(y|x,s)$ for any $(y,x,s)\in \cY \times \cX \times \cS$ and ergodic behavior so that the above capacity characterization is indeed valid.

In the following we summarize the results on error exponent upper bounds for unifilar channels derived in~\cite{AnWu17b}. The error exponent is upper bounded as $E(\overline{R}) \leq C_1(1-\overline{R}/C)$, where $\overline{R}$ is the average rate, $C$ is the channel capacity and $C_1 =  \sup_{s^0,b^1} \limsup_{N\rightarrow\infty}V^N(s^0,b^1)$, where $V^N(s^0,b^1)$ is the (average) reward in $N$ steps of a controlled Markov process with state $(S^0_t,B^1_{t-1})\in \cS \times \cP(\cS)$, action $(X^0_t,X^1_t) \in \cX \times (\cS \rightarrow \cP(\cX))$, instantaneous reward $R(S^0_t,B^1_{t-1},X^0_t,X^1_t)$ and transition kernel as given in~\cite[eq. (7)]{AnWu17b}.

An alternative upper bound on $C_1$ is given by $\tilde{C}_1 =   \max_{s^0,s^1} \limsup_{N\rightarrow\infty}V^N(s^0,s^1)$,
where $V^N(s^0,s^1)$ is the (average) reward in $N$ steps of a (simplified) controlled Markov process with state $(S^0_t,S^1_t)\in \cS^2$, action $(X^0_t,X^1_t) \in \cX^2$, instantaneous reward $\tilde{R}(S^0_t,S^1_t,X^0_t,X^1_t)=D(Q(\cdot|s^0,x^0)||Q(\cdot|s^1,x^1))$,
and transition kernel
\begin{align}
\label{eq:simplifiedMDP}
&\tilde{Q}(S^0_{t+1},S^1_{t+1}|S^0_{t},S^1_{t},X^0_{t},X^1_{t}) \nonumber \\
& = \sum_y \delta_{g(S^0_{t},X^0_t,y)}(S^0_{t+1})\delta_{g(S^1_{t},X^1_t,y)}(S^1_{t+1}) Q(y|X^0_t,S^0_{t}).
\end{align}

\begin{figure}[h]
\center
\includegraphics[width=0.4\columnwidth]{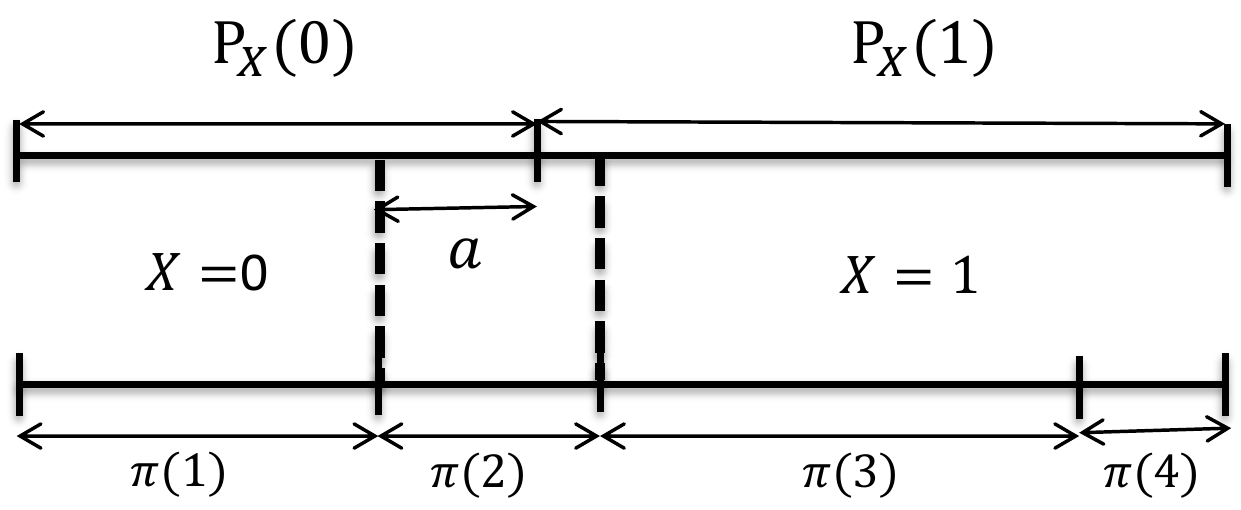}
\caption{Illustration of the ``discrete randomized posterior matching'' encoding.}
\label{fig:encodingstrategy}
\end{figure}

We now describe an encoding scheme that we will refer to extensively. This is exactly the scheme described in~\cite{Bu76} for DMCs.
For any message index $w\in\{1,\ldots,M\}\eqdef [M]$, input pmf $P_X\in\cP(\cX)$, message pmf $\pi\in\cP([M])$ and randomization variable $v\in[0,1]$ we define the ``discrete randomized posterior matching'' (DRPM) encoding\footnote{Although we use the term posterior matching, this is not to be confused with the encoding scheme described in~\cite{ShFe11} and uses probability density functions assuming messages take values in the continuoum $[0,1]$.} through the time-invariant mapping $X=pm(w,P_X,\pi,v)$, described as follows.
Partition the interval $[0,1]$ with subintervals of length $P_X(x)$ for $x\in \cX$.
Similarly, partition the interval $[0,1]$ with subintervals of length $\pi(i)$ for $i\in [M]$.
Transmit symbol $x$ with probability equal to the ratio of the length of the intersection of the intervals corresponding to $P_X(x)$ and $\pi(w)$ over $\pi(w)$.
This randomization is generated with the help of the random variable $v$ in an arbitrary (but prescribed) fashion.
Figure~\ref{fig:encodingstrategy} illustrates this encoding for $|\cX|=2$ and $M=4$. In this case, $X=0$ for $w=1$,
$X=1$ for $w=3,4$, and for $w=2$ the output $X$ depends on the value of $v$, i.e., if $v<a$, then $X=0$, otherwise $X=1$.

\section{Transmission scheme}
\label{sec:transmission}

In this section, we adapt Burnashev's scheme~\cite{Bu76} to unifilar channels.
To introduce randomness into our encoding strategy, in addition to the structure described in section~\ref{sec:model}, the transmitter and the receiver have access to a common set of random variables $\{V_t\}_{t \geq 1}$ in a causal way as described below.

\subsection{First stage}
Let $W\in[M]$ (with $M=2^K$) be the message to be transmitted and $P_e$ the target error probability. In this system, the transmitter receives perfect feedback of the output with unit delay and determines the input $X_t= e_t(W,Y^{t-1},V_t,S_1)$ with (deterministic) encoding strategies $\{e_t\}_{t\geq 1}$ at time $t$.
We emphasize that the encoding strategies are deterministic since this is a crucial part of the analysis: it allows the receiver to have an estimate of the current state and input for any hypothesized message. The common random variables are utilized to induce the appropriate maximizing input distribution.
Define the filtration $\{\mathcal{F}_t = (Y^{t},V^{t},S_1)\}_{t\geq1}$ and define the posterior probability of the massage, $\Pi_{t}$ as
\begin{equation}
\Pi_{t}(i) \eqdef P(W=i|\mathcal{F}_{t}) \qquad \forall i \in [M].
\end{equation}
We now specify in more detail the encoding strategy. Suppose we are given the collection of input distributions $\{P_X(\cdot|s,b)\in \cP(\cX)\}_{s,b\in \cS\times \cP(\cS)}$ that maximize~\eqref{eq:cap_singleletter}.
We define the vector $\underline{S}_t=(S^i_t)_{i=1}^M$, where $S^i_t$ is the state at time $t$ conditioning on $W=i$.
By recursive application of the function $g$ and the encoding functions $e_t$, state $S_t^i$ can be written as $S^i_t=g_t(i,Y^{t-1},V^{t-1},S_1)$.
We further define for any $s\in \cS$ and $i\in [M]$ the quantities
\begin{subequations}
\begin{align}
\label{eq:B_hat}
\hat{B}_{t-1}(s) &\eqdef  \sum_{i=1}^{M}\Pi_{t-1}(i)\delta_{S^i_t}(s)  \\
\label{eqn:conditionalbeliefonS}
\Pi^s_{t-1}(i) &\eqdef \frac{\Pi_{t-1}(i)\delta_{S^i_t}(s)}{\hat{B}_{t-1}(s)},
\end{align}
\end{subequations}
which are almost surely equal to $P(S_t=s|\mathcal{F}_{t-1})$ and $P(W=i|S_t=s,\mathcal{F}_{t-1})$, respectively.
The common random variables $\{V_t=(V_t^1,\cdots,V_t^{|\cS|})\in [0,1]^{|\cS|}\}_{t\geq 1}$ are generated as
\begin{equation}
P(V_t|V^{t-1},X^{t-1},S^t,Y^{t-1},W) = \prod_{i=1}^{|\cS|} u(V_t^i),
\end{equation}
where $u(\cdot)$ denotes the uniform distribution.
The transmission scheme is a generalization of Burnashev's scheme~\cite{Bu76} and also similar to the PM scheme~\cite{ShFe11}. First, based on the quantity $\Pi_{t-1}, \underline{S}_t, \hat{B}_{t-1}$, and $S^W_t=S_t$ the conditional distribution on the message given the state is evaluated as in~\eqref{eqn:conditionalbeliefonS}. Then the input signal $X_t$ is generated exactly as in DMC from $\Pi^{S_t^W}_{t-1}$ to ``match" the input distribution $P_X(\cdot|S_t,\hat{B}_{t-1})$, i.e.,
%
%
\begin{subequations}
\begin{align}
X_t&=pm(W,P_{X|SB}(\cdot|S^W_t,\hat{B}_{t-1}),\Pi^{S^W_{t}}_{t-1},V^{S^W_t}_t) \\
 &= e(W,\underline{S}_t,\Pi_{t-1},V_t^{S^W_t}).
\end{align}
\end{subequations}

The quantities $\ve{S}_{t+1}$ and $\Pi_t$ can be updated as
\begin{subequations}
\begin{align}
\Pi_t(i)&= \frac{Q(Y_t|e(i,\underline{S}_t,\Pi_{t-1},V_t^{S_t^i}),S^i_t)\Pi_{t-1}(i)}{\sum_j Q(Y_t|e(j,\underline{S}_t,\Pi_{t-1},V_t^{S_t^j}),S^j_t)\Pi_{t-1}(j)} \\
S^i_{t+1} &= g(S^i_{t},e(i,\underline{S}_t,\Pi_{t-1},V^{S^i_t}_t),Y_t),
\end{align}
\end{subequations}
which can be concisely written as
$\Pi_{t} = \phi_\pi(\underline{S}_t,\Pi_{t-1},Y_t,V_t)$,
$\underline{S}_{t+1} = \phi_s(\underline{S}_t,\Pi_{t-1},Y_t,V_t)$.
%
%
This process continues until $\max_i \Pi_{t-1}(i)$ exceeds a pre-specified threshold $p_0$.
This implies the receiver has very high confidence that a certain message is the transmitted one.
At this point, the transmitter enters the second stage that helps resolve whether the estimated message of the receiver is the true message, which is a binary hypothesis testing problem.

\subsection{Second stage}

Suppose we are given the optimizing strategies relating to the MDP discussed in~\cite{AnWu17b}.
In particular we are given a policy $X^0:\cS\times \cP(\cS) \rightarrow \cX$ and a policy
$X^1:\cS\times \cP(\cS) \rightarrow (\cS\rightarrow \cP(\cX))$.
At the end of stage one we have a message estimate $\hat{W}_{t}=\arg\max_i \Pi_{t-1}(i)$.
Let $H_0$ be the hypothesis that the estimation at the receiver is correct (i.e., $W=\hat{W}_t$), and $H_1$ be the opposite.
We define the quantities $\hat{B}^1_{t-1} \in \cP(\cS)$ and
$\Pi^1_{t-1} \in \cP([M])^{|\cS|}$ similarly to stage-one related quantities, with the only difference being that they represent posterior beliefs conditioned on $H_1$
\begin{subequations}
\begin{align}
\hat{B}^1_{t-1}(s) &= P(S_{t}=s|\mathcal{F}_{t-1},H_1)  \label{def:2stagebhat}\nonumber \\
&= \frac{\sum_{i\neq \hat{W}_t}\Pi_{t-1}(i)1_{\{S^i_{t}=s\}}}{1-\Pi_{t-1}(\hat{W}_t)} \\
\label{def:2stagepi1}
\Pi^{1,s}_{t-1}(i) &= P(W=i|S_t=s,\mathcal{F}_{t-1},H_1)\nonumber \\
 &=\frac{\Pi_{t-1}(i)1_{\{i\neq \hat{W}_t\}}1_{\{S^i_{t}=s\}}}{\hat{B}^1_{t-1}(s)(1-\Pi_{t-1}(\hat{W}_t))}.
\end{align}
\end{subequations}
Under $H_0$, the transmitted signal is $X_t = X^0[S^{\hat{W}_t}_t,\hat{B}^1_{t-1}]$. Under $H_1$, the transmitted signal $X_t$ is generated in a similar fashion as in stage one, i.e., using a DRPM scheme, expect that now, the message distribution $\Pi^{1,S^W_t}_{t-1}(\cdot)$ is used (instead of $\Pi^{S_t^W}_{t-1}(\cdot)$) and the input distribution $X^1[S^{\hat{W}_t}_t,\hat{B}^1_{t-1}](\cdot|S^W_t)$ is to be ``matched" (instead of $P_{X|SB}(\cdot|S_t,\hat{B}_{t-1})$).
We use $e^0(S^{\hat{W}_t}_t,\hat{B}^1_{t-1})$ and $e^1(S^{\hat{W}_t}_t,\hat{B}^1_{t-1},W,S^W_t,V_t)$ to denote the encoding functions for $H_0$ and $H_1$ respectively, where we make explicit the dependence on the common random variable $V_t$ used in $H_1$.
\begin{subequations}
\begin{align}
H_0: \  X_t&= X^0[S^{\hat{W}_t}_t,\hat{B}^1_{t-1}] = e^0(S^{\hat{W}_t}_t,\hat{B}^1_{t-1})\\
H_1: \  X_t&=pm(W,X^1[S^{\hat{W}_t}_t,\hat{B}^1_{t-1}](\cdot|S^W_t),\Pi^{1,S^W_t}_{t-1},V^{S^W_t}_t) \nonumber \\
 &= e^1(S^{\hat{W}_t}_t,\hat{B}^1_{t-1},W,S^W_t,V_t).
\end{align}
\end{subequations}
With this encoding strategy, we can update $\underline{S}_t$ and $\Pi_{t}$ by
\begin{subequations}
\begin{align}
S^{\hat{W}_t}_{t+1 } &= g(S^{\hat{W}_t}_{t },e^0(S^{\hat{W}_t}_{t},\hat{B}^1_{t-1}),Y_{t})  \\
S^{i}_{t+1 } &= g(S^{i}_{t },e^1(S^{\hat{W}_t}_{t},\hat{B}^1_{t-1},i,S^i_t,V_t),Y_{t}),  \quad  i\neq \hat{W}_t
\end{align}
and
\begin{align}
\Pi_t(\hat{W}_t)
 & = \frac{Q(Y_t|e^0(S^{\hat{W}_t}_{t},\hat{B}^1_{t-1}),S^{\hat{W}_t}_{t })\Pi_{t-1}(\hat{W}_t)}{P(Y_t|V_t,\cF_{t-1})} \\
 \Pi_t(i) & = \frac{Q(Y_t|e^1(S^{\hat{W}_t}_{t},\hat{B}^1_{t-1},i,S^i_t,V_t),S^i_t)\Pi_{t-1}(i)}{P(Y_t|V_t,\cF_{t-1})},
\quad  i\neq \hat{W}_t,
\end{align}
where the denominator is given by
\begin{align}
P(Y_t&|V_t,\cF_{t-1})
 = Q(Y_t|e^0(S^{\hat{W}_t}_{t},\hat{B}^1_{t-1}),S^{\hat{W}_t}_{t })\Pi_{t-1}(\hat{W}_t) \nonumber \\
&+\sum_{i\neq \hat{W}_t}Q(Y_t|e^1(S^{\hat{W}_t}_{t},\hat{B}^1_{t-1},i,S^i_t,V_t),S^i_t)\Pi_{t-1}(i).
\end{align}
\end{subequations}
Note that if during stage two the posterior belief of message $\hat{W}_t$ drops below the threshold $p_0$ then the system reverts to stage one.
Finally, we define a stopping time $T_{max}$ which is the first time that $\max_i\Pi_{t}(i)$ is greater than $1-P_e$, where $P_e$ is the pre-specified target error probability.
At this time transmission stops and the decoded message is declared to be $\hat{W}_{T_{max}}$.

\subsection{Alternative second stage}

We now describe an alternative, simplified, second stage for the case where the unifilar channel has the property that for every $s\in\cS$ there exists an $x=X^*(s)\in \cX$ s.t. $g(s,x,y)=g(s,X^*(s),y)=g'(y)$. Such channels are, for instance, those with $g(s,x,y)=s\oplus x \oplus y$, where $X^*(s)=s$.

Suppose we are given the function $X^*$ and the optimizing strategies relating to the simplified MDP discussed in~\eqref{eq:simplifiedMDP}.
In particular we are given policies $X^0:\cS^2 \rightarrow \cX$ and $X^1:\cS^2 \rightarrow \cX$.
The purpose of the first transmitted symbol at stage two is to perfectly inform the receiver about the state $S_{t+1}$. This is done by transmitting $X_t=X^*(S_t)$.
The remaining transmissions at stage two are as follows.
Under $H_i$, the transmitted signal and state update are given by
\begin{subequations}
\begin{align}
H_i: \qquad  X_t &= X^i[S^0_t,S^1_t], \quad i=0,1\\
S^i_{t+1} &= g(S^i_t,X^i[S^0_t,S^1_t],Y_{t}), \quad i=0,1.
\end{align}
\end{subequations}
In addition, the posterior belief is updated according to
\begin{subequations}
\begin{align}
\Pi_t(\hat{W}_t)
 & = \frac{Q(Y_t|X^0[S^0_t,S^1_t],S^0_t)\Pi_{t-1}(\hat{W}_t)}{P(Y_t|V_t,\cF_{t-1})}  \\
 \Pi_t(i) & = \frac{Q(Y_t|X^1[S^0_t,S^1_t],S^1_t)\Pi_{t-1}(i)}{P(Y_t|V_t,\cF_{t-1})},
\quad  i\neq \hat{W}_t,
\end{align}
where the denominator is given by
\begin{align}
P(Y_t&|V_t,\cF_{t-1})
 = Q(Y_t|X^0[S^0_t,S^1_t],S^0_t)\Pi_{t-1}(\hat{W}_t) \nonumber \\
&+Q(Y_t|X^1[S^0_t,S^1_t],S^1_t)(1-\Pi_{t-1}(\hat{W}_t)).
\end{align}
\end{subequations}

We conclude this section by pointing out that a simpler version of this transmission scheme is one that operates always in stage one, completely forgoing stage two, and stops once the maximum posterior message belief has crossed the threshold $1-P_e$.

\section{Error analysis}
\label{sec:analysis}

The average transmission rate of this system is defined as $\overline{R} = \frac{K}{\EE[T]}$.
The drift analysis provided below shows that $T_{max}$ is almost surely finite and therefore $P(err) < P_e$.
Inspired by Burnashev's method, we analyze the log-likelihood ratio $L_t$  at time $t$,  defined by
\begin{equation}
L_t = \log \frac{\Pi_t(W)}{1-\Pi_t(W)},
\end{equation}
and stopping time $T$ given by
\begin{equation}
T = \min \{t| L_t>\log \frac{1-P_e}{P_e}\}.
\end{equation}
%
To derive a lower bound on error-exponent, we would like to have an upper bound on $T_{max}$. In the following we derive an upper bound on $T$ instead of $T_{max}$ since $T$ is almost surely greater than $T_{max}$ and directly related to the message $W$. We first analyze the drift of $L_t$ w.r.t. $\{\mathcal{F}_t\}_{t\geq 0}$ and then apply the optional sampling theorem to a proposed submartingale, in order to derive the desired result. A major difference between unifilar channels and DMCs is the presence of memory. In order to capture channel memory, we eventually need to analyze multi-step instead of one-step drift of $L_t$.
\begin{lemma}  \label{lemma:onestepdrift}
During the first transmission stage the one-step drift of $\{L_t\}_{t\geq 0}$ w.r.t. $\{\mathcal{F}_t\}_{t\geq 0}$ is lower-bounded by
\begin{equation}
\EE[L_t-L_{t-1}|\mathcal{F}_{t-1}] \geq   I(\hat{B}_{t-1}),
\end{equation}
where $I(\hat{B}_{t-1})$ is given by
\begin{align}
I&(\hat{B}_{t-1})  = \sum_{s,x,y} Q(y|x,s)P_{X|SB}(x|s,\hat{B}_{t-1})\hat{B}_{t-1}(s) \nonumber \\
 &\log  \frac{Q(y|x,s)}{\sum_{\tilde{s},\tilde{x}} Q(y|\tilde{x},\tilde{s})P_{X|SB}(\tilde{x}|\tilde{s},\hat{B}_{t-1})\hat{B}_{t-1}(\tilde{s})}.
\end{align}
\end{lemma}
\begin{IEEEproof}
Please see Appendix~\ref{proof:onestepdrift}.
\end{IEEEproof}
The following remark is crucial in the subsequent development.
%
First, unlike the case in DMC where the one-step drift is shown to be greater than the channel capacity, here the one step drift is a random variable. This is exactly the reason we consider multi-step drift analysis: under an ergodicity assumption, the arithmetic mean of these random variables will converge almost surely to their mean.
This raises the question of what the mean of the process $\{I(\hat{B}_{t-1})\}_{t}$ is.
If the process $\{\hat{B}_{t-1}\}_t$ had the same statistics as those of the Markov chain $\{B_{t-1}\}_t$ defined in~\eqref{eq:B_MC}, then convergence to $C$ would be guaranteed with rate independent of  the parameter $K$. However, the two processes have different statistics. This is because of the introduction of common randomness! Indeed, $B_{t-1}(s)=P(S_{t}=s|Y^{t-1},S_1)$, while $\hat{B}_{t-1}(s)=P(S_{t}=s|Y^{t-1},V^{t-1},S_1)$ and they are related according to $B_{t-1}=\EE[\hat{B}_{t-1}|Y^{t-1},S_1]$. In fact, $\{\hat{B}_{t-1}\}_t$  is not a Markov chain, but is measurable w.r.t. the state of the Markov chain $\{(\ve{S_t},\Pi_{t-1})\}_t$ as shown in~\eqref{eq:B_hat}.
The following Lemma shows that during stage one, and in particular when $\max_i\Pi_{t-1}(i)<\epsilon$, the process $\{\hat{B}_{t-1}\}_t$ has approximately the same statistics as the process $\{B_{t-1}\}_t$.
\begin{lemma}  \label{lemma:approxBhat}
If $\max_i\Pi_{t-1}(i)<\epsilon$ then $||P_{\hat{B}|\cF}(\cdot|\cF_{t-1}) - P_{B|B}(\cdot|\hat{B}_{t-1})|| = o_{\epsilon}(1)$ in an appropriately defined metric (such as the Wasserstein metric).
\end{lemma}
\begin{IEEEproof}
Please see Appendix~\ref{proof:approxBhat}.
\end{IEEEproof}
The intuition behind this result is that during the time when all posterior message beliefs are small, the effect of randomization is negligible and thus the random variables $V^{t-1}$ do not reveal any additional information about the state $S_t$ than is revealed by $Y^{t-1},S_1$.
Based on the above lemma, and employing a continuity argument we expect that the process $\{\hat{B}_{t-1}\}_t$
converges arbitrarily close to the steady state distribution $\pi_B$ independently of $K$. This in turn implies from Lemma~\ref{lemma:onestepdrift} that the quantity $\frac{1}{N}\sum_{t=1}^N I(\hat{B}_{t-1})$ approaches $C$
for sufficiently large $N$, during the time when $\max_i\Pi_{t-1}(i)<\epsilon$.

We now analyze the drift of $\{L_t\}_t$ during stage two. In particular we consider the ``alternative'' second stage and channels with the special property described therein.
The analysis follows the same structure as in the first stage: we first analyze the one-step drift of $L_t$ under both hypotheses and then, arguing in the same way we did above, we consider the asymptotic behaviour of the $N$-step drift.

\begin{lemma} \label{lemma:twostagemultistepdrift}
For any $\epsilon>0$, there exist an $N=N(\epsilon)$ such that
\begin{subequations}
\begin{align}
\EE[L_{t+N}-L_t|\mathcal{F}_t] \geq N(\tilde{C}_1-\epsilon) \quad  \text{if } L_t  \geq \log\frac{p_0}{1-p_0}\\
\EE[L_{t+N}-L_t|\mathcal{F}_t] \geq N(C-\epsilon) \quad  \text{if } L_t  < \log\frac{p_0}{1-p_0}.
\end{align}
\end{subequations}
\end{lemma}
\begin{IEEEproof}
See Appendix~\ref{proof:twostagemultistepdrift}.
\end{IEEEproof}
Lemma~\ref{lemma:twostagemultistepdrift} shows that in the second stage, the likelihood ratio grows faster than in the first stage if the estimation at the receiver is correct. Even if the estimation is wrong, the likelihood ratio maintains the increasing rate as in the first stage.
For this to be true we have assumed that a related quantity $\tilde{C}_1^*>\tilde{C}$. If this is not the case then an alternative scheme can be proposed as in~\cite[p.~261]{Bu76}. We omit this description due to space limitations.

Furthermore, due to the assumption that the transition kernel $Q(y|x,s)$ positive for any $(s,x,y)\in \cS\times \cX \times \cY$, the quantity $|L_t-L_{t-1}|$ can be shown to be upper bounded by a certain constant $C_2$ as in~\cite[Lemma~4]{Bu76}.

Collecting the drift results for stage one and ``alternative'' stage two
and utilizing the connection between the drift and the stopping time as in~\cite[Lemma, p.50]{burnashev1975one},
we can derive a lower bound on the error exponent.
The result is given in the form of a conjecture due to the issues raised after Lemma~\ref{lemma:onestepdrift}.
\begin{conjecture} \label{thm:twostagescheme}
With $M=2^K$ messages and target error probability $P_e$, given any $\epsilon >0$ we have
\begin{equation}
-\frac{\log P_e}{\EE[T]} \geq \tilde{C}_1(1-\frac{\overline{R}}{C}) + U(\epsilon,K,P_e,C,\tilde{C}_1,C_2),
\end{equation}
where $\lim_{K\rightarrow \infty }U(\epsilon,K,P_e,C,\tilde{C}_1,C_2)=o_\epsilon(1)$.
\end{conjecture}
\begin{IEEEproof}
See Appendix~\ref{proof:twostagescheme}.
\end{IEEEproof}

\section{Numerical results}
\label{sec:example}

In this section, we provide simulation results for the error exponents achieved by the two proposed transmission schemes for some binary input/output/state unifilar channels.
We consider the trapdoor channel (denoted as channel $A$), chemical channel (denoted as channel $B(p_0)$), symmetric unifilar channel (denoted as $C(p_0,q_0)$, and asymmetric unifilar channel (denoted as $D(p_0,q_0,p_1,q_1)$.
All of these channels have $g(s,x,y) = s\oplus x \oplus y$ and kernel $Q$ as shown in Table~\ref{t:Q}.
\begin{table}[H]
\centering
\caption{Kernel definition for binary unifilar channels}
\begin{tabular}{|c|c|c|c|c|}
\hline
Channel & $Q(0| 0,0)$& $Q(0|1,0)$ & $Q(0|0,1)$&  $Q(0|1,1)$\\
\hline
Trapdoor & 1 & 0.5 & 0.5 & 0\\
\hline
Chemical($p_0$) & $1$ & $p_0$ & $1-p_0$ & 0\\
\hline
Symmetric($p_0,q_0$) & $1-q_0$ & $p_0$ & $1-p_0$ & $q_0$\\
\hline
Asymmetric($p_0,q_0,p_1,q_1$) & $1-q_0$ & $p_0$ & $1-p_1$ & $q_1$\\
\hline
\end{tabular}
\label{t:Q}
\end{table}
We simulated a system with message length $K=10,20,30$ (bits) and target error rates $P_e=10^{-3}, 10^{-6}, 10^{-9}, 10^{-12}$. In each simulation sufficient experiments were run to have a convergent average rate, since the error probability is guaranteed to be bellow the target. Infinite precision arithmetic was used in all evaluations through the ``GNU Multiple Precision Arithmetic Library" (GMP).

\begin{figure}[htbp]
\centering
\includegraphics[width=0.5\columnwidth]{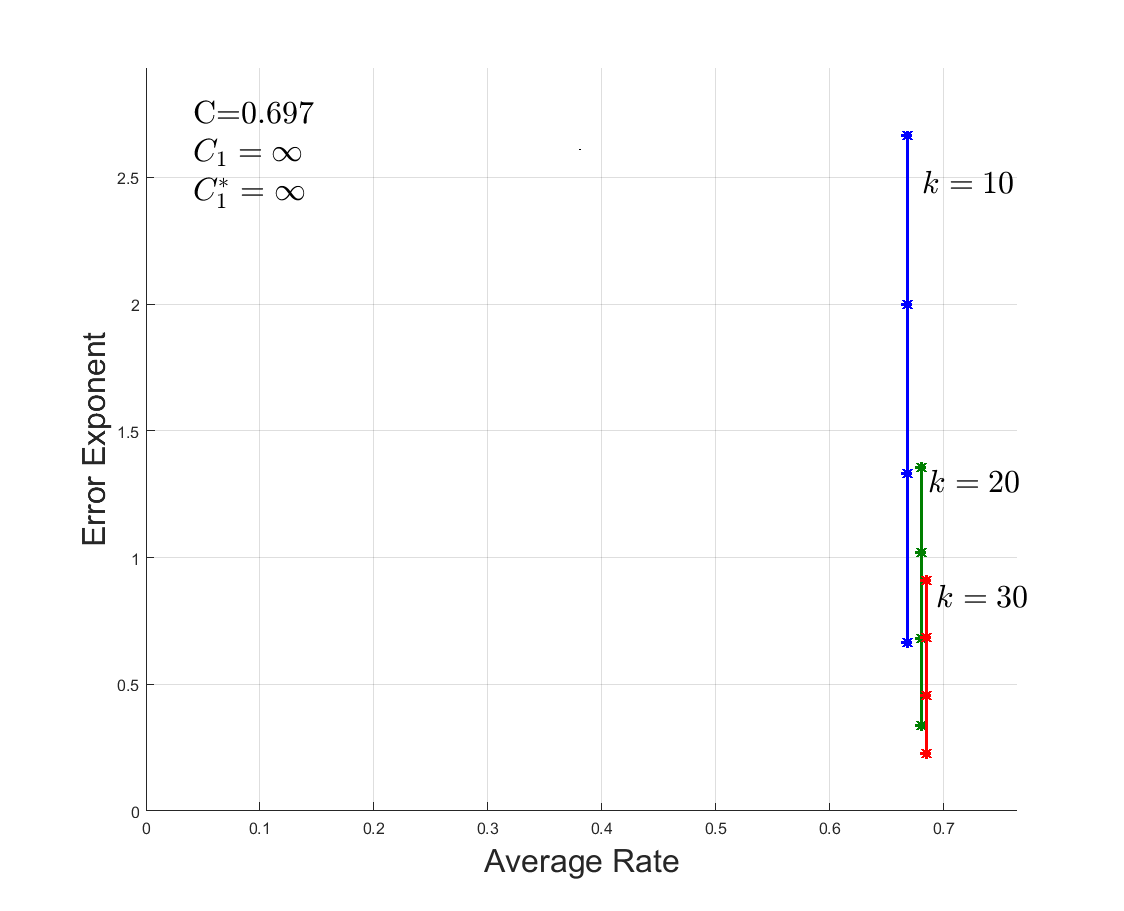}
\caption{Error exponent for the trapdoor channel A.}
\label{fig:chemical1}
\end{figure}

\begin{figure}[htbp]
\centering
\includegraphics[width=0.5\columnwidth]{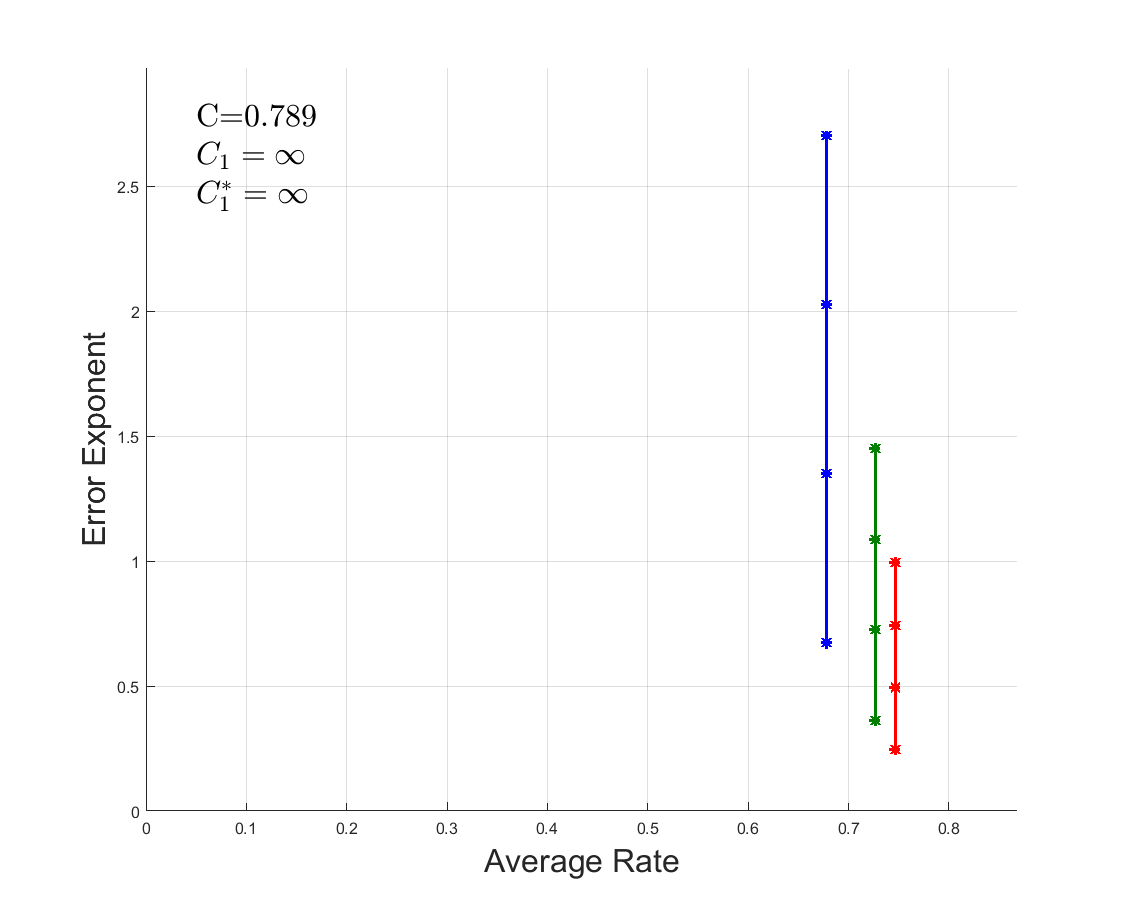}
\caption{Error exponent for the chemical channel B(0.9).}
\label{fig:chemical2}
\end{figure}

\begin{figure}[htbp]
\centering
\includegraphics[width=0.5\columnwidth]{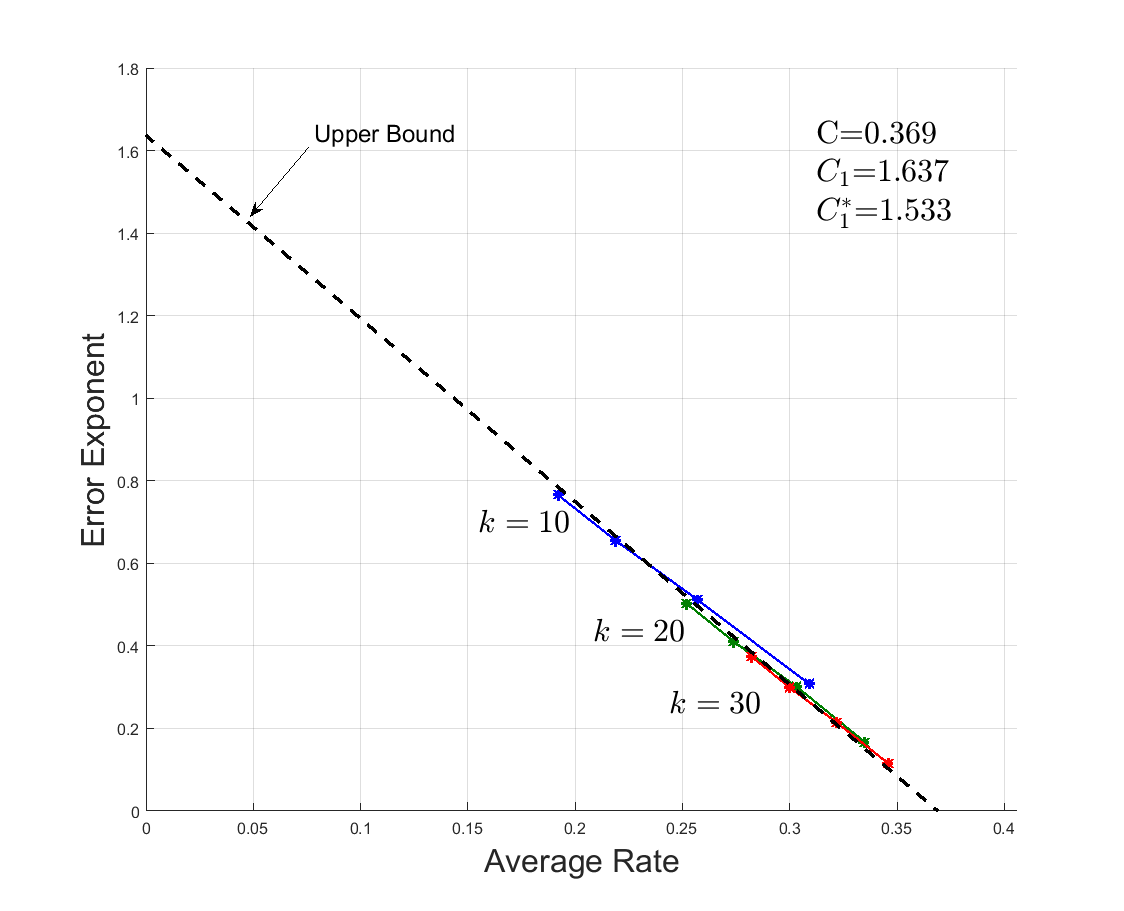}
\caption{Error exponent for the symmetric unifilar channel C(0.5,0.1).}
\label{fig:symmetric1}
\end{figure}

\begin{figure}[htbp]
\centering
\includegraphics[width=0.5\columnwidth]{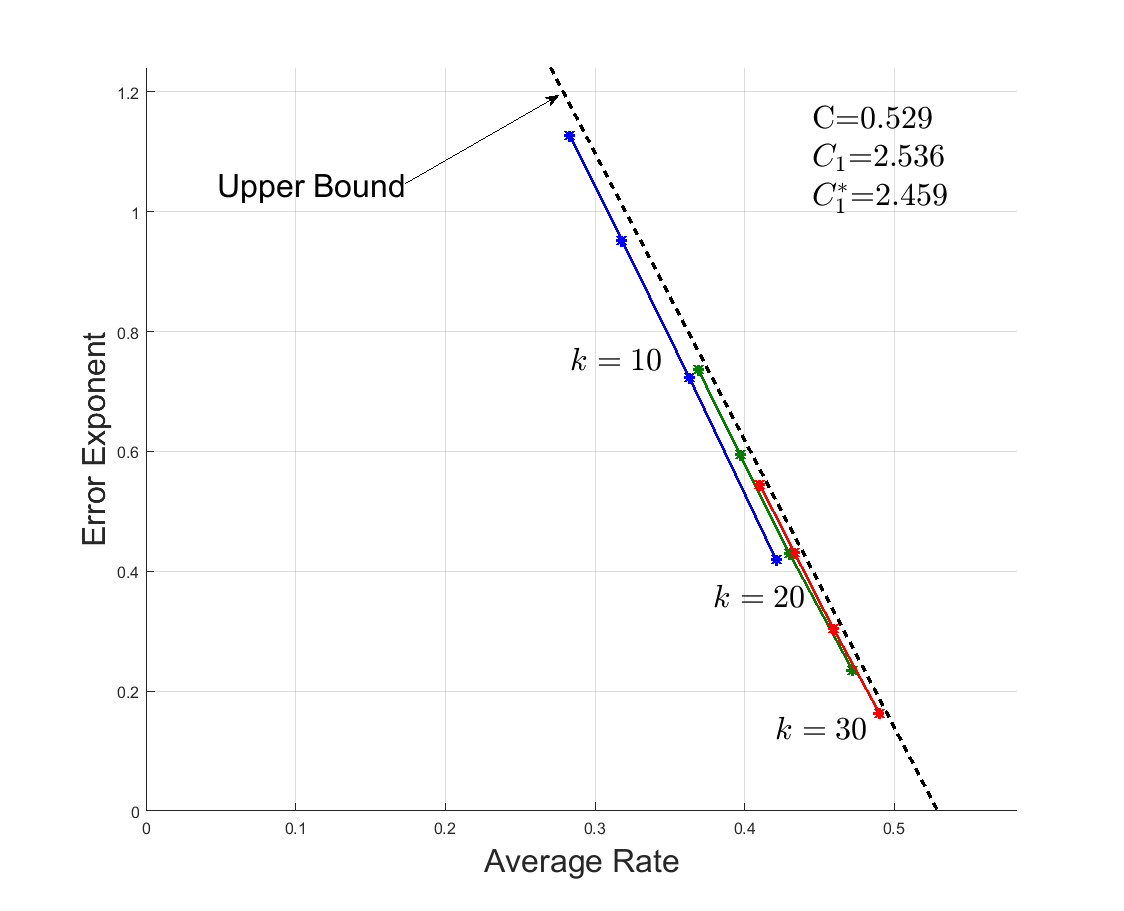}
\caption{Error exponent for the symmetric unifilar channel C(0.9,0.1).}
\label{fig:symmetric2}
\end{figure}

\begin{figure}[htbp]
\centering
\includegraphics[width=0.5\columnwidth]{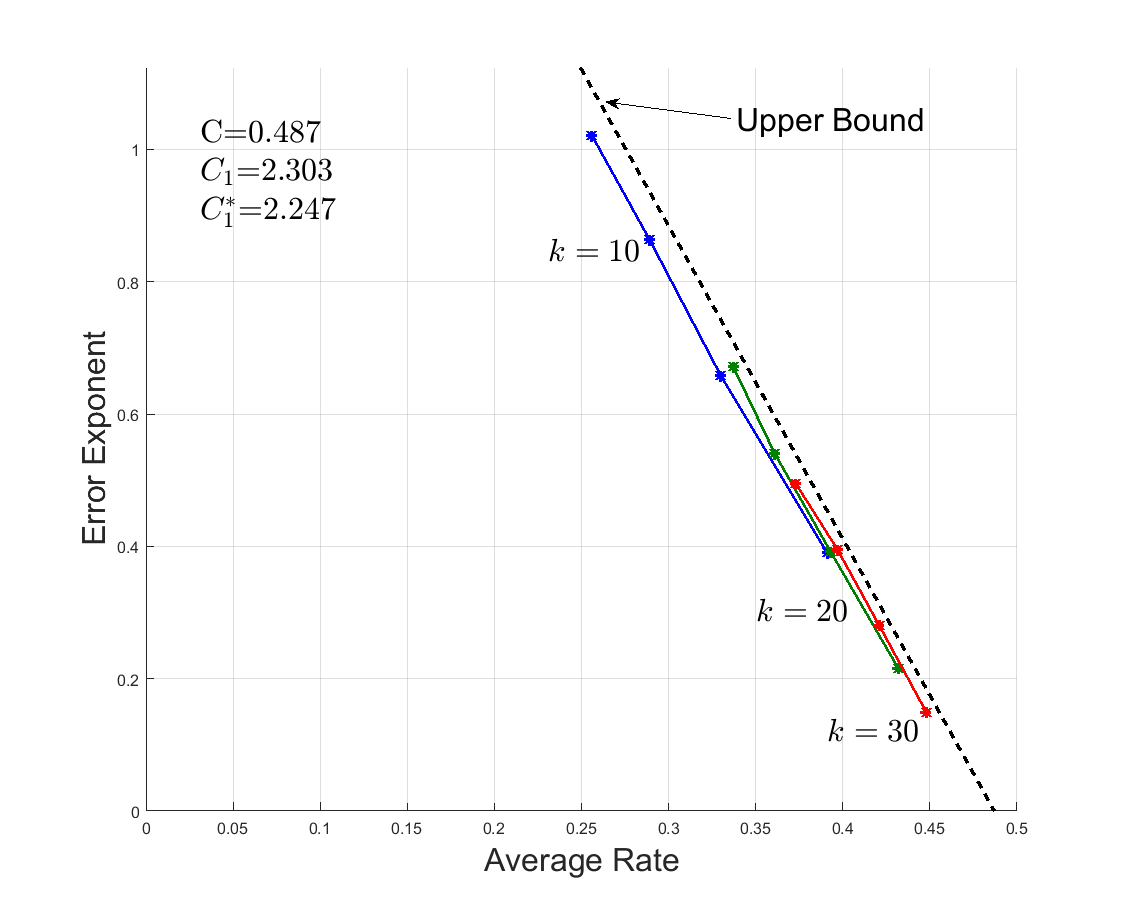}
\caption{Error exponent for the asymmetric unifilar channel D(0.5,0.1,0.1,0.1).}
\label{fig:asymmetric1}
\end{figure}

\begin{figure}[htbp]
\centering
\includegraphics[width=0.5\columnwidth]{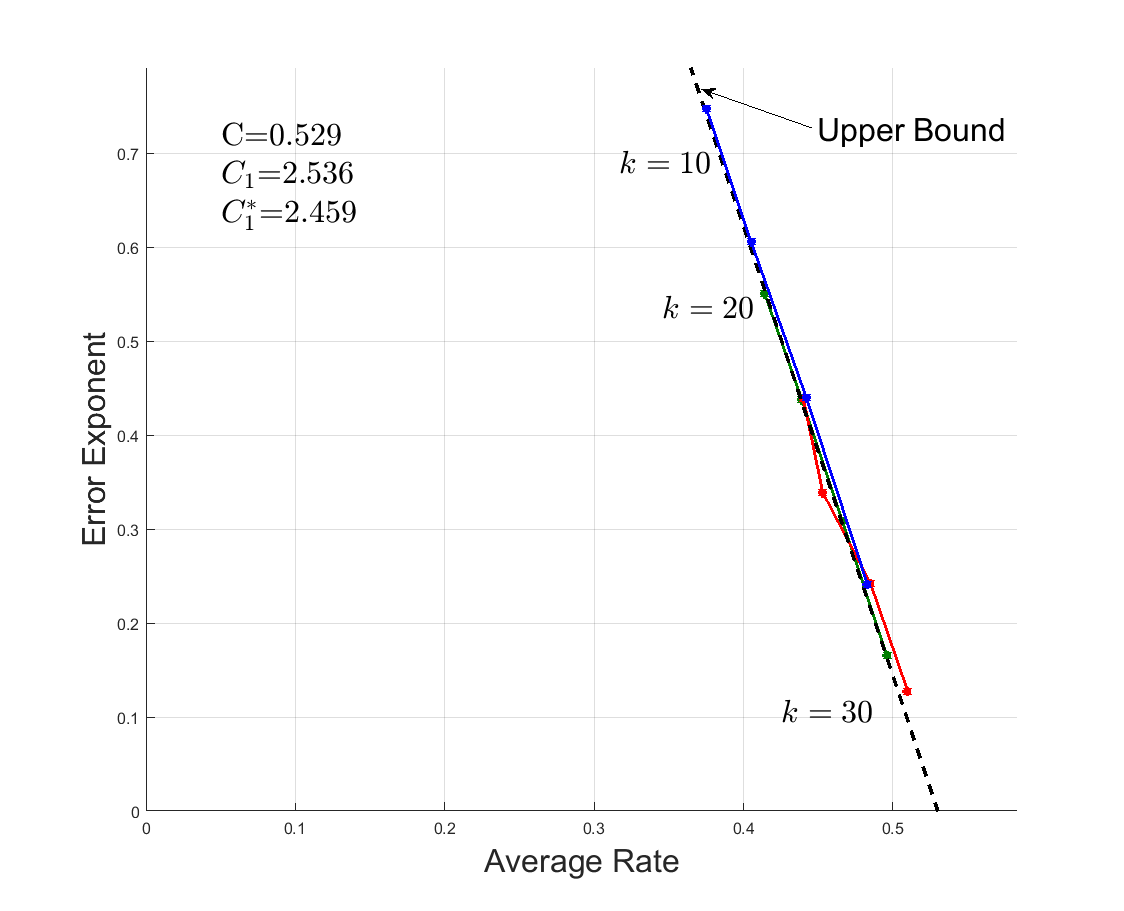}
\caption{Error exponent for the asymmetric unifilar channel D(0.9,0.1,0.1,0.1).}
\label{fig:asymmetric2}
\end{figure}

The results are shown in Figs.~\ref{fig:chemical1}--\ref{fig:asymmetric2}. Each curve in these figures corresponds to a value of $K$.
All two-stage schemes were run with the more elaborate optimal policy (i.e., not the ``alternative'' one).
Also shown on the same figures are the error exponent upper bounds, as well as the parameters $(C,C_1,C_1^*)$ for each channel.



We make two main observations regarding these results.
The first observation is that for the trapdoor channels there is strong evidence that the error exponents are infinite.
This is consistent with the findings in~\cite{PeCuVaWe08} where a zero-error capacity-achieving scheme is proposed, and the discussion in~\cite{AnWu17b} regarding channels with zeros in their transition kernels.
Similar comments are valid for the chemical channel.
The second observation is the remarkable agreement between simulation results of the proposed (two-stage) scheme and the upper bound derived in~\cite{AnWu17b} for
channels C and D.
%
These results represent very strong evidence for the validity of the conjecture stated earlier.
%

\section{Conclusions}
\label{sec:conclusions}
We propose a variable-length transmission scheme for unifilar channels with noiseless feedback and analyze its error exponent by generalizing the techniques of Burnashev~\cite{Bu76}.
The techniques used in this work can also be applied to channels with Markov states and inter-symbol interference (ISI) where the state is observed at the receiver and with unit delay at the transmitter.

\clearpage
\appendices

\section{Proof of Lemma~\ref{lemma:onestepdrift}}
\label{proof:onestepdrift}

\begin{align}
E&[L_t-L_{t-1}|W,Y^{t-1},V^{t-1},S_1]\nonumber \\
 &= \log(1-\Pi_{t-1}(W)) + \EE[\log \frac{\Pi_t(W)}{(1-\Pi_t(W))\Pi_{t-1}(W)} |W,\mathcal{F}_{t-1}]\nonumber \\
 &=  \log(1-\Pi_{t-1}(W)) + \EE[\log \frac{Q(Y_t|X_t,S_t)}{P(Y_t|V_t,\mathcal{F}_{t-1})-Q(Y_t|X_t,S_t)\Pi_{t-1}(W)} |W,\mathcal{F}_{t-1}].
\end{align}
We look into $P(Y_t=y|V_t,\mathcal{F}_{t-1})$
\begin{align}
P(Y_t=y|V_t,\mathcal{F}_{t-1}) &= \sum_{x,s} Q(y|x,s) \sum_{i=1}^M P(X_t =x, S_t=s, W=i|V_t,\mathcal{F}_{t-1})\nonumber \\
&= \sum_{x,s} Q(y|x,s) \sum_{i=1}^M \delta_{e(i,\underline{S}_t,\Pi_{t-1},V^{S^i_t}_t)}(x)\delta_{S^i_t}(s)\Pi_{t-1}(i),
\end{align}
and after taking expectations
\begin{align}
E&[P(Y_t=y|V_t,\mathcal{F}_{t-1})|W,X_t,S_t ,\mathcal{F}_{t-1}]\nonumber \\
 &= \EE[\sum_{x,s} Q(y|x,s) \sum_{i=1}^M \delta_{e(i,\underline{S}_t,\Pi_{t-1},V^{S^i_t}_t)}(x)\delta_{S^i_t}(s)\Pi_{t-1}(i)|W,X_t,S_t ,V^{t-1},Y^{t-1},S_1] \nonumber \\
&= \sum_{x,s} Q(y|x,s) \sum_{i=1}^M \EE[\delta_{e(i,\underline{S}_t,\Pi_{t-1},V^{S^i_t}_t)}(x)|W,X_t,S_t ,\mathcal{F}_{t-1}]\delta_{S^i_t}(s)\Pi_{t-1}(i) \nonumber \\
&= \sum_{x,s} Q(y|x,s) \sum_{i=1}^M \overline{e}(x|i,\underline{S}_t,\Pi_{t-1})\delta_{S^i_t}(s)\Pi_{t-1}(i)  \nonumber \\
&= \sum_{x,s} Q(y|x,s) (P_X(x|s,\hat{B}_{t-1})+\delta_{X_t}(x)\delta_{S_t}(s)\Pi_{t-1}(W)-\overline{e}(x|W,\underline{S}_t,\Pi_{t-1})\delta_{S_t}(s)\Pi_{t-1}(W))\nonumber \\
&= P(Y_t=y | \hat{B}_{t-1}) + Q(y|X_t,S_t)\Pi_{t-1}(W) - \Pi_{t-1}(W)\sum_{x}Q(y|x,S_t)\overline{e}(x|W,\underline{S}_t,\Pi_{t-1}),
\end{align}
where $\overline{e}(x|i,\underline{S}_t,\Pi_{t-1})$ is given by
\begin{align}
\overline{e}(x|i,\underline{S}_t,\Pi_{t-1})=\EE[\delta_{e(i,\underline{S}_t,\Pi_{t-1},V^{S^i_t}_t)}(x)|W,X_t,S_t ,V^{t-1},Y^{t-1},S_1].
\end{align}
Therefore, by convexity of $f(x) = \log\frac{A}{x-B}$,
\begin{align} \label{ineq:onestepdrift1}
E&[L_t-L_{t-1}|W,\mathcal{F}_{t-1}] \nonumber \\
&\geq \log(1-\Pi_{t-1}(W))
+  \EE[\log \frac{Q(Y_t|X_t,S_t)}{P(Y_t|\hat{B}_{t-1})-\Pi_{t-1}(W)\sum_{x}Q(Y_t|x,S_t)\overline{e}(x|W,\Pi_{t-1},S_t)} |W,\mathcal{F}_{t-1} ] .
\end{align}
Looking further into the last term in the above inequality we get
\begin{align}  \label{ineq:onestepdrift2}
E&[\log  \frac{Q(Y_t|X_t,S_t)}{P(Y_t|\hat{B}_{t-1})-\Pi_{t-1}(W)\sum_{x}Q(Y_t|x,S_t)\overline{e}(x|W,\Pi_{t-1},S_t)} |W,\mathcal{F}_{t-1} ] \nonumber \\
&=E [\log  \frac{Q(Y_t|X_t,S_t)}{P(Y_t|\hat{B}_{t-1})} +\log\frac{P(Y_t|\hat{B}_{t-1})}{P(Y_t|\hat{B}_{t-1})-\Pi_{t-1}(W)\sum_{x}Q(Y_t|x,S_t)\overline{e}(x|W,\Pi_{t-1},S_t)} |W,\mathcal{F}_{t-1} ] \nonumber \\
&\geq E [\log  \frac{Q(Y_t|X_t,S_t)}{P(Y_t|\hat{B}_{t-1})}|W,\mathcal{F}_{t-1} ] + \EE[\log\frac{1}{1-\Pi_{t-1}(W)}|W,\mathcal{F}_{t-1} ],
\end{align}
where the last equation is due to the convexity of $x\log\frac{1}{1-A x}$. Combining \eqref{ineq:onestepdrift1} and \eqref{ineq:onestepdrift2}, we get
\begin{align}
\EE[L_t-L_{t-1}|\mathcal{F}_{t-1}] &= \EE[\EE[L_t-L_{t-1}|W,\mathcal{F}_{t-1}]|\mathcal{F}_{t-1}]\nonumber \\
&\geq \EE[\log  \frac{Q(Y_t|X_t,S_t)}{P(Y_t|\hat{B}_{t-1})}| \mathcal{F}_{t-1}] \nonumber \\
&= I(\hat{B}_{t-1}).
\end{align}

\section{Proof of Lemma~\ref{lemma:approxBhat}}
\label{proof:approxBhat}

Consider the update equation for $\hat{B}_{t}$
\begin{subequations}
\begin{align}
\hat{B}_t(s)&\eqdef P(S_{t+1}=s|\cF_t) \\
 &=\frac{\sum_{x,s'}  P(S_{t+1}=s,X_t=x,S_t=s',Y_t,V_t|\cF_{t-1}) }{\sum_{s,x,s'}  P(S_{t+1}=s,X_t=x,S_t=s',Y_t,V_t|\cF_{t-1})} \\
 &=\frac{\sum_{x,s'}  \delta_{g(s',x,Y_t)}(s) Q(Y_t|x,s') P(X_t=x|S_t=s',V_t,\cF_{t-1}) P_V(V_t) \hat{B}_{t-1}(s')}{\sum_{x,s'}  Q(Y_t|x,s') P(X_t=x|S_t=s',V_t,\cF_{t-1}) P_V(V_t) \hat{B}_{t-1}(s') } \\
 &=\frac{\sum_{x,s'}  \delta_{g(s',x,Y_t)}(s) Q(Y_t|x,s') P(X_t=x|S_t=s',V_t,\cF_{t-1}) \hat{B}_{t-1}(s')}{\sum_{x,s'}  Q(Y_t|x,s') P(X_t=x|S_t=s',V_t,\cF_{t-1})  \hat{B}_{t-1}(s') }
\end{align}
\end{subequations}
Consider the quantity $P(X_t=x|S_t=s',V_t,\cF_{t-1})$. We have
\begin{subequations}
\begin{align}
P&(X_t=x|S_t=s',V_t,\cF_{t-1}) = \nonumber \\
 &= \sum_{i=1}^M P(X_t=x,W=i|S_t=s',V_t,\cF_{t-1}) \\
 &= \sum_{i=1}^M P(X_t=x|W=i,S_t=s',V_t,\cF_{t-1}) P(W=i|S_t=s',V_t,\cF_{t-1}) \\
 &= \sum_{i=1}^M \delta_{e(i,\underline{S}_t,\Pi_{t-1},V_t^{s'})}(x) \Pi^{s'}_{t-1}(i) \\
 &= \sum_{i<i_0} \delta_{0}(x) \Pi^{s'}_{t-1}(i) + \sum_{i>i_0} \delta_{1}(x) \Pi^{s'}_{t-1}(i) \nonumber  \\
 &\quad + \delta_{0}(x) 1(V_t^{s'}\Pi^{s'}_{t-1}(i_0)<a) \Pi^{s'}_{t-1}(i)
        + \delta_{1}(x) 1(V_t^{s'}\Pi^{s'}_{t-1}(i_0)>a) \Pi^{s'}_{t-1}(i) \\
 &= \sum_{i<i_0} \delta_{0}(x) \Pi^{s'}_{t-1}(i) + \sum_{i>i_0} \delta_{1}(x) \Pi^{s'}_{t-1}(i) \nonumber\\
 &\quad + \delta_{0}(x) 1(V_t^{s'}\Pi^{s'}_{t-1}(i_0)<a) \Pi^{s'}_{t-1}(i_0)
        + \delta_{1}(x) 1(V_t^{s'}\Pi^{s'}_{t-1}(i_0)>a) \Pi^{s'}_{t-1}(i_0) \nonumber\\
 &\quad + \delta_{0}(x) a + \delta_{1}(x) (\Pi^{s'}_{t-1}(i_0)-a)
        - \delta_{0}(x) a - \delta_{1}(x) (\Pi^{s'}_{t-1}(i_0)-a)  \\
 &\overset{(a)}{=}  P_{X|S,B}(x|s',\hat{B}_{t-1}) \nonumber\\
 &\quad + \delta_{0}(x) [1(V_t^{s'}\Pi^{s'}_{t-1}(i_0)<a) \Pi^{s'}_{t-1}(i_0)-a]
        + \delta_{1}(x) [a-1(V_t^{s'}\Pi^{s'}_{t-1}(i_0)<a) \Pi^{s'}_{t-1}(i_0)] \\
 &=  P_{X|S,B}(x|s',\hat{B}_{t-1}) + \Delta,
\end{align}
\end{subequations}
where for simplicity of notation we have considered the case of binary inputs, $a$ is as shown in Fig.~\ref{fig:encodingstrategy} and (a) is due to the PM encoding scheme.
Furthermore, $|\Delta|\leq \Pi^{s'}_{t-1}(i_0)\leq \epsilon / \hat{B}_{t-1}(s')$.
Substituting back to the update equation we get
\begin{subequations}
\begin{align}
\hat{B}_t(s)&\eqdef P(S_{t+1}=s|\cF_t) \\
 &=\frac{\sum_{x,s'}  \delta_{g(s',x,Y_t)}(s) Q(Y_t|x,s') [P_{X|S,B}(x|s',\hat{B}_{t-1}) + \Delta] \hat{B}_{t-1}(s')}{\sum_{x,s'}  Q(Y_t|x,s') [P_{X|S,B}(x|s',\hat{B}_{t-1}) + \Delta]  \hat{B}_{t-1}(s') } \\
 &=\frac{\sum_{x,s'}  \delta_{g(s',x,Y_t)}(s) Q(Y_t|x,s') P_{X|S,B}(x|s',\hat{B}_{t-1})  \hat{B}_{t-1}(s') + \sum_{x,s'}  \delta_{g(s',x,Y_t)}(s) Q(Y_t|x,s') \Delta \hat{B}_{t-1}(s')}{\sum_{x,s'}  Q(Y_t|x,s') P_{X|S,B}(x|s',\hat{B}_{t-1}) \hat{B}_{t-1}(s') + \sum_{x,s'}   Q(Y_t|x,s') \Delta \hat{B}_{t-1}(s') } \\
 &=\frac{\sum_{x,s'}  \delta_{g(s',x,Y_t)}(s) Q(Y_t|x,s') P_{X|S,B}(x|s',\hat{B}_{t-1}) \hat{B}_{t-1}(s')}{P_{Y|B}(Y_t|\hat{B}_{t-1}) } \nonumber \\
  &\quad + \frac{Num}{P_{Y|B}(Y_t|\hat{B}_{t-1}) [P_{Y|B}(Y_t|\hat{B}_{t-1})+ \sum_{x,s'}   Q(Y_t|x,s') \Delta \hat{B}_{t-1}(s')]},
\end{align}
\end{subequations}
where
\begin{align}
Num &= P_{Y|B}(Y_t|\hat{B}_{t-1})[\sum_{x,s'}  \delta_{g(s',x,Y_t)}(s) Q(Y_t|x,s') \Delta \hat{B}_{t-1}(s')] \nonumber \\
 &\quad -[\sum_{x,s'}  \delta_{g(s',x,Y_t)}(s) Q(Y_t|x,s') P_{X|S,B}(x|s',\hat{B}_{t-1})  \hat{B}_{t-1}(s')][\sum_{x,s'}   Q(Y_t|x,s') \Delta \hat{B}_{t-1}(s')].
\end{align}
Taking into account that $|\Delta \hat{B}_{t-1}(s')|\leq \epsilon$, the update equation for $\hat{B}_t$ is $\hat{B}_t=\phi(\hat{B}_{t-1},Y_t)+o_{\epsilon}(1)$, where $\phi(\cdot)$ is the update equation for the quantity $B_t$ in~\eqref{eq:B_rec}.

Following a similar reasoning we have for $P(Y_t|\cF_{t-1})$
\begin{subequations}
\begin{align}
P(Y_t|\cF_{t-1}) &=\sum_{x,s'}  Q(Y_t|x,s') P(X_t=x|S_t=s',\cF_{t-1}) \hat{B}_{t-1}(s') \\
 &=\sum_{x,s'}  Q(Y_t|x,s') P_{X|S,B}(x|s',\hat{B}_{t-1})\hat{B}_{t-1}(s') \\
 &=P_{Y|B}(Y_t|\hat{B}_{t-1}).
\end{align}
\end{subequations}
Combining the above we get
\begin{align}
P(\hat{B}_t|\cF_{t-1})
 =\sum_{y} \delta_{\phi(\hat{B}_{t-1},y)+o_{\epsilon}(1)}(\hat{B}_t) P_{Y|B}(y|\hat{B}_{t-1})
\end{align}
which can be compared with the kernel $P_{B|B}$
\begin{align}
P(B_t|Y^{t-1},S_1)
 =\sum_{y} \delta_{\phi(B_{t-1},y)}(B_t) P_{Y|B}(y|B_{t-1}).
\end{align}
Comparing the two kernels, it is clear that these are ``close'' in an appropriately defined distance
(such as the Wasserstein metric). This is evident by the fact that each of these distributions puts the same
probability mass to points differing by $o_{\epsilon}(1)$.

\section{Proof of Lemma~\ref{lemma:twostagemultistepdrift}}
\label{proof:twostagemultistepdrift}

\begin{align}
\label{eqn:onestepdriftfor2ndstage}
\EE[L_{t+1}-L_{t}|\mathcal{F}_{t},H_0]
 &=  \EE[\log\frac{\Pi_{t+1}(\hat{W}_t)}{1-\Pi_{t+1}(\hat{W}_t)}-\log \frac{\Pi_{t}(\hat{W}_t)}{1-\Pi_{t}(\hat{W}_t)}|\mathcal{F}_{t},H_0]\nonumber \\
 &= \EE[\log \frac{Q(Y_{t+1}|X^0[S^0_{t+1},S^1_{t+1}],S^0_{t+1})}{ Q(Y_{t+1}|X^1[S^0_{t+1},S^1_{t+1}],S^1_{t+1})}|\mathcal{F}_{t},H_0] \nonumber \\
 &=\tilde{R}(S^0_{t+1},S^1_{t+1},X^0[S^0_{t+1},S^1_{t+1}],X^1[S^0_{t+1},S^1_{t+1}]),
\end{align}
where  $\tilde{R}(s^0,s^1,x^0,x^1)$ is given by
\begin{align}
\tilde{R}(s^0,s^1,x^0,x^1) = \sum_y Q(y|x^0,s^0)\log \frac{Q(y|x^0,s^0)}{Q(y|x^1,s^1)}.
\end{align}
We can think of this quantity as the instantaneous reward received by the Markov chain  $(S^0_{t},S^1_{t})_{t\geq 1}$.
Considering the $N$-step drift, we have
\begin{align}
\EE[L_{t+N}-L_{t}|\mathcal{F}_{t},H_0] &= \sum_{i=t}^{t+N-1} \EE[L_{i+1}-L_i|\mathcal{F}_{t},H_0]\nonumber \\
&= \sum_{i=t}^{t+N-1} \EE[L_{i+1}-L_i|\mathcal{F}_{t},H_0] \nonumber \\
&= \sum_{i=t}^{t+N-1} \EE[ \EE[L_{i+1}-L_i|\mathcal{F}_{i},H_0  ]|\mathcal{F}_{t},H_0 ]\nonumber \\
&\overset{(a)}{=}  \sum_{i=t}^{t+N-1} \EE[\EE[\log \frac{Q(Y_{i+1}|X^0[S^0_{i+1},S^1_{i+1}],S^0_{i+1})}{Q(Y_{i+1}|X^1[S^0_{i+1},S^1_{i+1}],S^1_{i+1})}|\mathcal{F}_{i},H_0]|\mathcal{F}_{t},H_0 ] \nonumber \\
&= \sum_{i=t}^{t+N-1} \EE[  \tilde{R}(S^0_{i+1},S^1_{i+1},X^0[S^0_{i+1},S^1_{i+1}],X^1[S^0_{i+1},S^1_{i+1}] ) |\mathcal{F}_t,H_0],
\end{align}
where (a) is due to~\eqref{eqn:onestepdriftfor2ndstage}.
Thus, the multi-step drift corresponds to the total average reward in the aforementioned Markov chain. This total reward relates to the MDP problem discussed in~\cite{AnWu17b}.
Thus, under an ergodicity assumption,
the corresponding per-unit-time reward converges to $\tilde{C}_1$  almost surely as $N\rightarrow \infty$.
Thus, given any $\epsilon >0$ , there exist a $N_1= N_1(\epsilon)$  such that
\begin{equation}
\EE[L_{t+N_1}-L_{t}|F_t,H_0] \geq N_1(\tilde{C}_1-\epsilon).
\end{equation}

In the case when $L_t <  \log\frac{p_0}{1-p_0}$ then there are two cases.
If $W$ is still the most reliable message then we are still under hypothesis $H_0$ and the above analysis holds. If however, $H_1$ is true then
\begin{align} \label{eqn:mutistepdrift2ndstageH1}
\EE[&L_{t+1}-L_t|\cF_t,H_1] \nonumber \\
&= \EE[\log\frac{Q(Y_{t+1}|X^1[S^0_{t+1},S^1_{t+1}],S^1_{t+1})(1-\Pi_{t}(W))}{ \sum_{i\neq W,\hat{W}_{t}}  Q(Y_{t+1}|X^1[S^0_{t+1},S^1_{t+1}],S^1_{t+1})\Pi_t(i)+ Q(Y_{t+1}|X^0[S^0_{t+1},S^1_{t+1}],S^0_{t+1})\Pi_t(\hat{W}_{t})   }|\cF_t,H_1].
\end{align}
As $\Pi_{t}(\hat{W}_{t}) \rightarrow 1$, the right-hand side of~\eqref{eqn:mutistepdrift2ndstageH1} becomes
\begin{align}
\EE[\log\frac{Q(Y_{t+1}|X^1[S^0_{t+1},S^1_{t+1}],S^1_{t+1})}{Q(Y_{t+1}|X^0[S^0_{t+1},S^1_{t+1}],S^0_{t+1})}|\cF_t,H_1] = \tilde{R}^*_1(S^0_{t+1},S^1_{t+1},X^0[S^0_{t+1},S^1_{t+1}],X^1[S^0_{t+1},S^1_{t+1}]).
\end{align}

This quantity can be thought of as some instantaneous reward received by the Markov chain  $(S^0_{t},S^1_{t})_{t\geq 1}$ induced by the policies $X^0,X^1$.
Thus, under the ergodicity assumption stated earlier,
the corresponding per-unit-time reward converges to $\tilde{C}^*_1 \geq C$  almost surely as $N\rightarrow \infty$.
Therefore, under hypothesis $H_1$, for any $\epsilon>0$, there exists $N_2=N_2(\epsilon)$ such that
\begin{equation}
\EE[L_{t+N}-L_{t}|\cF_t,H_1]  \geq N_2(C-\epsilon).
\end{equation}
Selecting $N=\max\{N_1,N_2\}$ concludes the proof.

\section{Proof of Conjecture~\ref{thm:twostagescheme}}
\label{proof:twostagescheme}

Define a process $Z_t = L_{Nt}$ and filtration $\cF'_t = \mathcal{F}_{Nt}$ . We have
\begin{align}
\EE[Z_{t+1}- Z_t|\cF'_{t}] &\geq N(C-\epsilon) \qquad \text{(this is the conjectured behaviour)}\nonumber \\
\EE[Z_{t+1}- Z_t|\cF'_{t}] &\geq   N(\tilde{C}_1-\epsilon) \qquad \text{if } Z_t > \log\frac{1-p_0}{p_0} \nonumber \\
|Z_t - Z_{t-1}| &\leq  NC_2 .
\end{align}

Define a stopping time $\tilde{T}$ w.r.t. to $\{\cF'_t\}$ by
\begin{equation}
\tilde{T} = \min \{t| Z_t \geq \log\frac{1-P_e}{P_e} \}.
\end{equation}
By definition we have $T \leq N\tilde{T}$  almost surely.
Applying \cite[Lemma, p.~50]{burnashev1975one} to $\{Z_t\}_{t\geq0}$, we have an upper bound on the expectation of the stopping time $\tilde{T}$.
\begin{align}
\frac{\EE[T]}{N} \leq \EE[\tilde{T}] \leq \frac{\log\frac{1-P_e}{P_e}}{N(\tilde{C}_1-\epsilon)} + \frac{K}{N(C-\epsilon)}+D(NC,NC_1,NC_2).
\end{align}
Rearranging the above inequality, we get
\begin{align}
-\frac{\log P_e}{\EE[T]} \geq \tilde{C}_1(1-\frac{\overline{R}}{C})  -\frac{N(\tilde{C}_1-\epsilon)D(NC,N\tilde{C}_1,NC_2)}{K/\overline{R}} -\frac{\log(1-P_e)}{K/\overline{R}}
+ \epsilon (\frac{(C-\tilde{C}_1)\overline{R}}{C(C-\epsilon)}-1).
\end{align}



\begin{thebibliography}{10}
\providecommand{\url}[1]{#1}
\csname url@rmstyle\endcsname
\providecommand{\newblock}{\relax}
\providecommand{\bibinfo}[2]{#2}
\providecommand\BIBentrySTDinterwordspacing{\spaceskip=0pt\relax}
\providecommand\BIBentryALTinterwordstretchfactor{4}
\providecommand\BIBentryALTinterwordspacing{\spaceskip=\fontdimen2\font plus
\BIBentryALTinterwordstretchfactor\fontdimen3\font minus
  \fontdimen4\font\relax}
\providecommand\BIBforeignlanguage[2]{{%
\expandafter\ifx\csname l@#1\endcsname\relax
\typeout{** WARNING: IEEEtran.bst: No hyphenation pattern has been}%
\typeout{** loaded for the language `#1'. Using the pattern for}%
\typeout{** the default language instead.}%
\else
\language=\csname l@#1\endcsname
\fi
#2}}

\bibitem{Ho63}
M.~Horstein, ``Sequential transmission using noiseless feedback,'' \emph{{IEEE}
  Trans. Inform. Theory}, vol.~9, no.~3, pp. 136--143, Jul 1963.

\bibitem{ScKa66}
J.~Schalkwijk and T.~Kailath, ``A coding scheme for additive noise channels
  with feedback--{I}: No bandwidth constraint,'' \emph{{IEEE} Trans. Inform.
  Theory}, vol.~12, no.~2, pp. 172--182, Apr 1966.

\bibitem{Bu76}
M.~V. Burnashev, ``Data transmission over a discrete channel with feedback.
  {R}andom transmission time,'' \emph{Problemy Peredachi Informatsii}, vol.~12,
  no.~4, pp. 10--30, Oct.-Dec. 1976.

\bibitem{ShFe11}
O.~Shayevitz and M.~Feder, ``Optimal feedback communication via posterior
  matching,'' \emph{IEEE Trans.~Information Theory}, vol.~57, no.~3, pp.
  1186--1222, Mar. 2011.

\bibitem{YaIt79}
H.~Yamamoto and K.~Itoh, ``Asymptotic performance of a modified
  schalkwijk-barron scheme for channels with noiseless feedback (corresp.),''
  \emph{IEEE Transactions on Information Theory}, vol.~25, no.~6, pp. 729--733,
  Nov 1979.

\bibitem{viswanathan1999capacity}
H.~Viswanathan, ``Capacity of {M}arkov channels with receiver csi and delayed
  feedback,'' \emph{Information Theory, IEEE Transactions on}, vol.~45, no.~2,
  pp. 761--771, 1999.

\bibitem{chen2005capacity}
J.~Chen and T.~Berger, ``The capacity of finite-state {M}arkov channels with
  feedback,'' \emph{IEEE Transactions on Information Theory}, vol.~51, no.~3,
  pp. 780--798, 2005.

\bibitem{yang2005feedback}
S.~Yang, A.~Kav{\v{c}}i{\'c}, and S.~Tatikonda, ``Feedback capacity of
  finite-state machine channels,'' \emph{Information Theory, IEEE Transactions
  on}, vol.~51, no.~3, pp. 799--810, 2005.

\bibitem{PeCuVaWe08}
H.~Permuter, P.~Cuff, B.~V. Roy, and T.~Weissman, ``Capacity of the trapdoor
  channel with feedback,'' \emph{IEEE Trans.~Information Theory}, vol.~54,
  no.~7, pp. 3150--3165, July 2008.

\bibitem{TaMi09}
S.~Tatikonda and S.~Mitter, ``The capacity of channels with feedback,''
  \emph{IEEE Trans.~Information Theory}, vol.~55, no.~1, pp. 323--349, Jan.
  2009.

\bibitem{WuAn16b}
J.~Wu and A.~Anastasopoulos, ``On the capacity of the chemical channel with
  feedback,'' in \emph{Proc.~International Symposium on Information Theory},
  Barcelona, Spain, July 2016.

\bibitem{StChKu16}
P.~A. Stavrou, C.~D. Charalambous, and C.~K. Kourtellaris, ``Sequential
  necessary and sufficient conditions for optimal channel input distributions
  of channels with memory and feedback,'' in \emph{2016 IEEE International
  Symposium on Information Theory (ISIT)}, July 2016, pp. 300--304.

\bibitem{BaAn10}
J.~H. Bae and A.~Anastasopoulos, ``A posterior matching scheme for finite-state
  channels with feedback,'' in \emph{Proc.~International Symposium on
  Information Theory}, Austin, TX, June 2010, pp. 2338--2342.

\bibitem{An12a}
A.~Anastasopoulos, ``A sequential transmission scheme for unifilar finite-state
  channels with feedback based on posterior matching,'' in
  \emph{Proc.~International Symposium on Information Theory}, July 2012, pp.
  2914--2918.

\bibitem{CoYuTa09}
G.~Como, S.~Yuksel, and S.~Tatikonda, ``The error exponent of variable-length
  codes over {M}arkov channels with feedback,'' \emph{IEEE Trans.~Information
  Theory}, vol.~55, no.~5, pp. 2139--2160, May 2009.

\bibitem{AnWu17b}
A.~Anastasopoulos and J.~Wu, ``Variable-length codes for channels with memory
  and feedback: error exponent upper bounds,'' Tech. Rep., Jan. 2017,
  (available online on arxiv at \url{https://arxiv.org/abs/1701.06678}).

\bibitem{burnashev1975one}
M.~V. Burnashev and K.~Zigangirov, ``On one problem of observation control,''
  \emph{Problemy Peredachi Informatsii}, vol.~11, no.~3, pp. 44--52, 1975.

\end{thebibliography}

\end{document}